\documentclass[11pt,twoside]{article}
\usepackage{hepro5}
\usepackage{graphicx}

\usepackage[T1]{fontenc} 

\usepackage{latexsym}
\usepackage{verbatim}
\usepackage{amssymb}

\begin{document}

\vskip 1.0cm
\markboth{A.T. Araudo et al.}{Magnetic field amplification in hotspots}
\pagestyle{myheadings}

\vspace*{0.5cm}
\title{Particle acceleration and magnetic field amplification in hotspots of
FR~II galaxies: The case study 4C74.26}

\author{A.T. Araudo$^1$, A.R. Bell$^2$, K.M.~Blundell$^1$}
\affil{$^1$University of Oxford, Astrophysics, Keble Road, 
Oxford OX1~3RH, UK\\
$^{2}$University of Oxford, Clarendon Laboratory, Parks Road, 
Oxford OX1~3PU, UK 
}

\begin{abstract}
It has been suggested that relativistic  shocks in extragalactic
sources may accelerate the most energetic cosmic rays. However, recent 
theoretical advances indicating that relativistic shocks are probably 
unable to 
accelerate particles to energies much larger than 1~PeV cast doubt on this.
In the present contribution we model the radio to X-ray emission in the 
southern hotspot of the  
quasar 4C74.26. The synchrotron radio emission is resolved near
the shock with the MERLIN radio-interferometer, and the rapid decay of  
this emission behind the shock is interpreted as the decay of the
downstream magnetic field as expected for small scale turbulence.  
If our result is confirmed by analyses of other
radiogalaxies, it provides firm observational evidence 
that relativistic shocks at the termination region of powerful jets
in FR II radiogalaxies do not accelerate ultra high
energy cosmic rays.
\end{abstract}

\section{Introduction}

Radiogalaxies are the subclass of Active Galactic Nuclei (AGN)
where jets are clearly detected 
at radio frequencies, which in turn are classified in type I and
II Faranoff-Riley (FR) galaxies. Hotspots are 
usually detected at the jet termination region of FR~II radiogalaxies.
These bright  radio synchrotron knots have a size $\sim 1-10$~kpc
and are embedded in larger lobes of shocked plasma. 
The location of the hotspot is coincident with the downstream region of the 
jet reverse shock, where particles accelerated by the latter emit
non-thermal radiation.

Diffusive shock acceleration (DSA) is a well established mechanism 
to accelerate
particles in astrophysical sources where shock waves are present
(Bell 1978a,b). Particles diffuse  back and forth across
the shock  and gain energy in each crossing.
Therefore, long times are required to accelerate the most energetic cosmic 
rays unless  the magnetic field around the shock is amplified.
The amplified turbulent field 
scatters particles rapidly so that they cross the shock more frequently 
achieving a higher energy in the available time.  
The state-of-the-art of DSA in the hotspots of FR~II radiogalaxies is a 
phenomenological picture where the acceleration process finishes 
when particles start to radiate their energy or when they can escape from
the source, i.e. the energy gained is sufficient for  the Larmor radius 
to exceed the size of the acceleration region (Meisenheimer et al. 1989). 
Assuming 
that the magnetic field persists over long distances 
downstream of the shock, the distribution of non-thermal emitting
electrons is a broken power-law where the break frequency
is determined by a competition between synchrotron losses and adiabatic
expansion (e.g. Brunetti et al. 2003).
We will see however that the magnetic field must 
be highly discrete to explain the very thin radio emission 
in the southern hotspot of the quasar 4C74.26 (Araudo et al. 2015).

\begin{figure} 
\begin{center}
\includegraphics[angle=0,height=7.0cm]{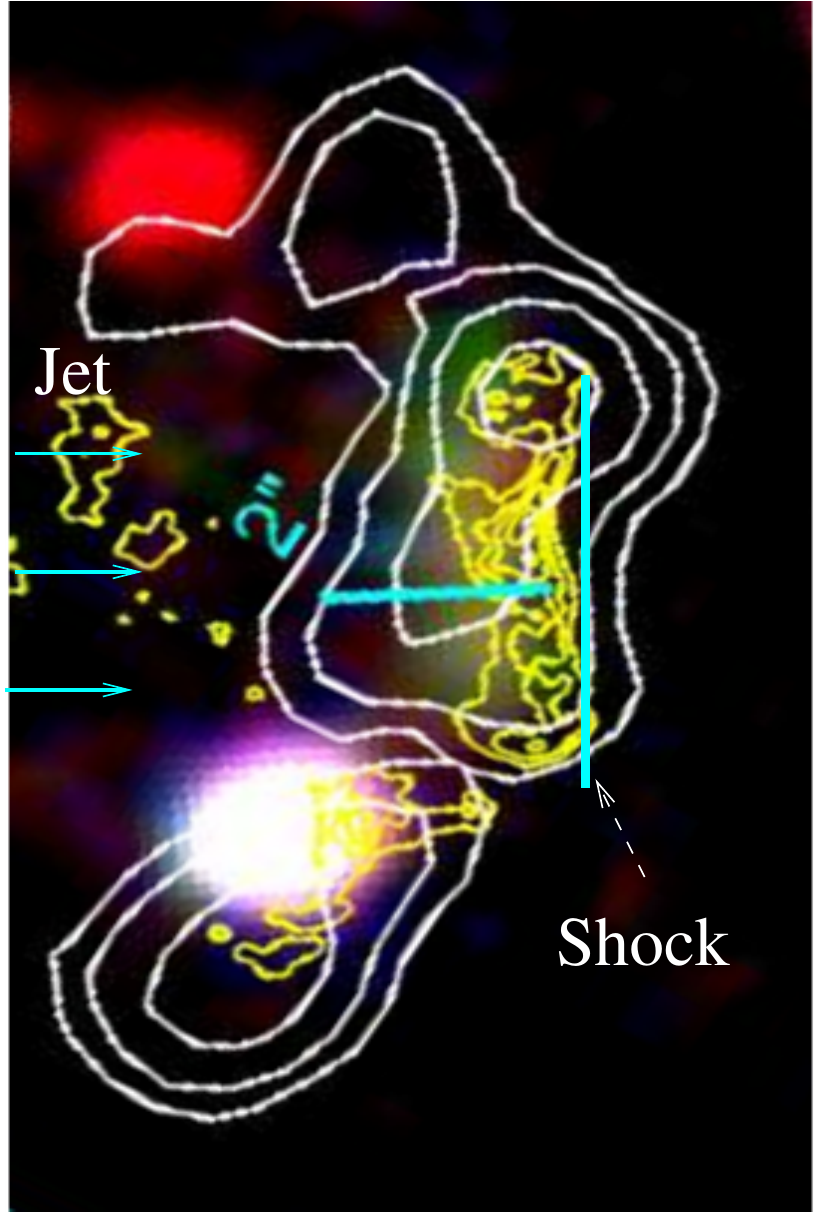}
\hspace*{0.2cm}
\includegraphics[height=7.0cm]{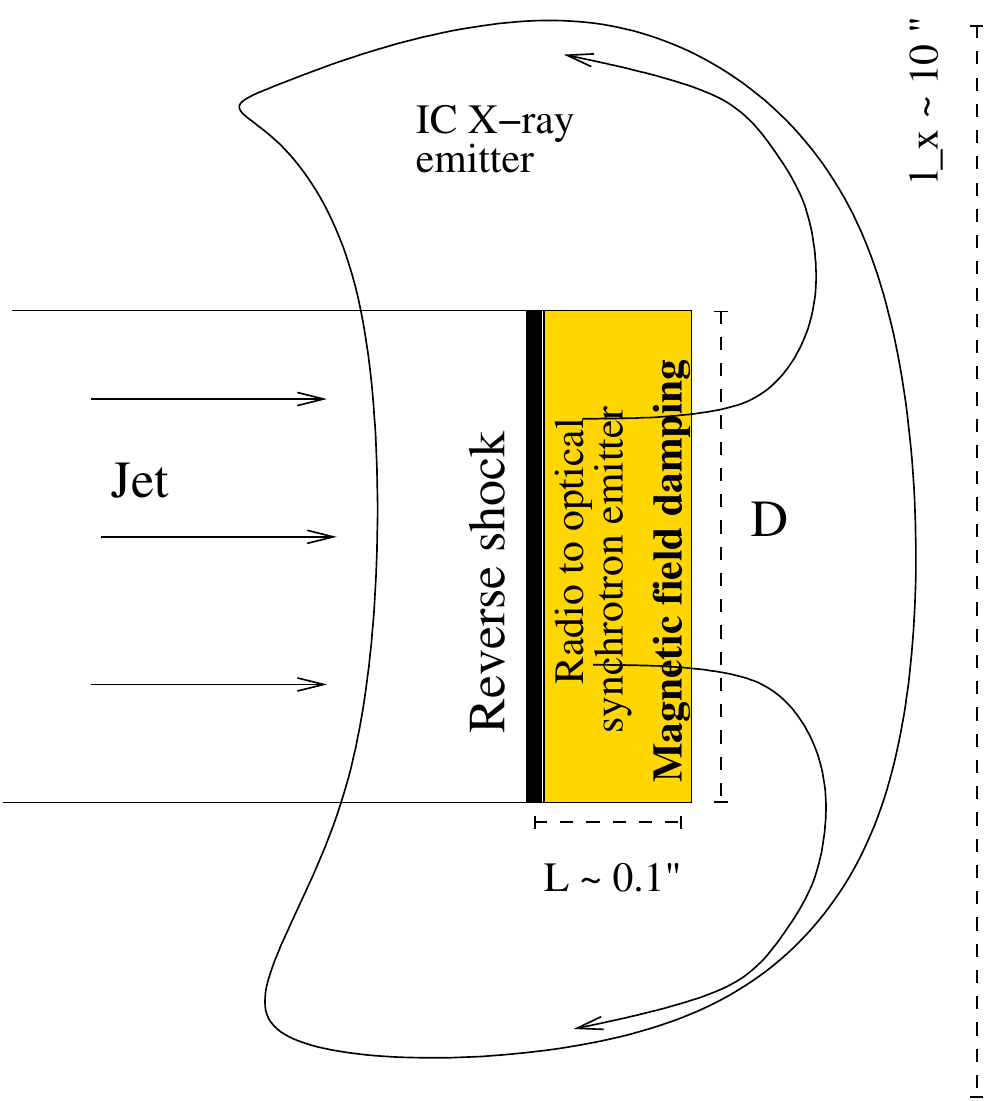}
\caption{\emph{Left:} Southern arc adapted from Erlund et al. (2010).
\emph{Right:} Sketch of the jet termination region (as seen at $\theta_{\rm j} = 90^{\circ}$). Particles are accelerated
at the reverse shock and radiate in the shock downstream region. 
}
\label{sketch}
\end{center}
\end{figure}

\section{The case study 4C74.26: the southern hotspot}

The FR~II radiogalaxy 4C74.26 is located at redshift $z = 0.104$ 
($\sim$0.5~Gpc from Earth)\footnote{Throughout the paper we use cgs units 
and the cosmology $H_0 = 71$~km~s$^{-1}$~Mpc$^{-1}$, $\Omega_0 = 1$ and 
$\Lambda_0 = 0.73$. One arcsecond represents
$1.887$~kpc on the plane of the sky at $z = 0.104$.}. This source is 
the largest known radio quasar, with a  projected linear size of 1.1~Mpc and
it has a one-sided jet.  
Two X-ray sources
have been detected with the \emph{Chandra} satellite in the termination region 
of the South jet (Erlund et al. 2007). The brighest
X-ray source is called the ``X-ray peak'' and has no counterpart
at other frequencies. On the other hand, the southern X-ray 
source is coincident with a radio, IR and optical source (Erlund et al. 2010).
In the present contribution we focus on this multiwavelength hotspot and 
assume that this is the jet termination region.  

The hotspot X-ray specific
luminosity is $L_{\rm x}\sim$7.9$\times10^{40}$~erg~s$^{-1}$ at frequency
$\nu_{\rm x} =2.4\times10^{17}$~Hz (2~keV).
The shape of this emission is arc-like with a characteristic size
$l_{\rm x} \sim 10^{\prime\prime}$, and 
encloses a compact radio source, also arc-like and called 
the ``southern arc'', as we show in Figure~\ref{sketch} (left). 
Compact radio emission from the southern arc was detected 
with the MERLIN high resolution interferometer at a frequency 
$\nu_{\rm r} = 1.66$~GHz with specific luminosity 
$L_{\rm r} \sim$ 1.9$\times$10$^{40}$~erg~s$^{-1}$. 
This emission is located in a   
region of width $l_{\rm r} < 1^{\prime\prime}$ on the plane of the sky.
In addition, faint and diffuse radiation was 
detected at IR ($\nu_{\rm ir} =$ 1.36$\times$10$^{14}$~Hz) and optical 
($\nu_{\rm opt} =$ 6.3$\times$10$^{14}$~Hz) bands, 
and located in a region of width $\gtrsim l_{\rm r}$. 
However, there is a linear structure in both bands that traces the 
brightest edge
of the MERLIN radio emission, and seems to be cupped within it.

The measured radio-to-IR spectral index is $\alpha = 0.75$, typical of 
synchrotron radiation, and the steep spectrum between IR and optical 
indicates that the cut-off of the emission is at 
$\nu_{\rm ir} \le\nu_{\rm c} \le \nu_{\rm opt}$ (see Fig.~13 in Erlund et al. 2010). 
Therefore, the southern arc 
X-ray emission is not synchrotron\footnote{Given that the southern hotspot 
is located at $\sim 0.5$~Mpc from the nucleus,  absorption of the 
emission by photoionization is ruled-out (see e.g. Ryter et al. 1996).}.

\subsection{Inverse Compton X-ray emission}

The X-ray emission from hotspots is usually  explained as synchrotron
self Compton (SSC) or up-scattering of Cosmic Microwave Background (CMB)
photons (e.g. Hardcastle et al. 2004, 
Werner et al. 2012). In the case of the southern arc in 4C74.26, SSC is 
disfavoured because 1) there is an off-set between the peak of
 the X-ray and radio emission, and 2) the energy density of synchrotron
photons is 8$\times$10$^{-15}$~erg~cm$^{-3}$, much smaller than the energy
density of CMB photons $U_{\rm cmb} = $6$\times$10$^{-13}$~erg~cm$^{-3}$
(Erlund et al. 2010). Therefore,  we consider that the X-ray emission from the 
southern arc  is produced by inverse Compton (IC) scattering of CMB radiation.

CMB photons with energy $E_{\rm cmb} \sim$7$\times10^{-4}$~eV are scattered
up to $2$~keV X-rays by electrons with Lorentz factor 
$\gamma_{\rm x} \sim \sqrt{h \nu_{\rm x}/E_{\rm cmb}}\sim 10^3$, where $h$ is the 
Planck constant.
Unless the macroscopic Lorentz factor of the jet is $\gtrsim$10, 
$\gamma_{\rm x}$-electrons are 
non-thermal and follow a power-law energy distribution
$N_e = K_e\gamma^{-p}$, with $p = 2\alpha +1 = 2.5$. 
The X-ray specific luminosity can be written as
$L_{\rm x} \sim N_e(\gamma_{\rm x}) \gamma_{\rm x}^3/t_{\rm ic}(\gamma_{\rm x}) V_{\rm x}$, where $t_{\rm ic}$ is the IC cooling timescale and $V_{\rm x}$ is the 
volume of the X-ray emitter. From the previous equation we estimate $K_e$ and
therefore, the electron energy density required to explain $L_{\rm x}$ is 
\begin{equation}
U_e \sim K_e \left(\frac{\gamma_{\rm min}^{2-p}}{p-2}\right) V_{\rm x} \sim 10^{-9} 
\left(\frac{\gamma_{\rm min}}{50}\right)^{-0.5}
\left(\frac{V_{\rm x}}{300\,{\rm arcsec}^3}\right)^{-1}\,\,{\rm erg\, cm}^{-3}, 
\end{equation}
where $N_e$ terminates at $\gamma_{\rm min}$.
The magnetic field in equipartition with non-thermal particles,
i.e. $B_{\rm eq}^2/(8\pi) = (1+a)U_e$, where $a$ takes into account the 
contribution of non-thermal protons, is
\begin{equation}
B_{\rm eq} \sim 160 (1+a)^{0.5}
\left(\frac{\gamma_{\rm min}}{50}\right)^{-0.25}
\left(\frac{V_{\rm x}}{300\,{\rm arcsec}^3}\right)^{-0.5}\,\,{\rm \mu G}.
\end{equation}
Note that $B_{\rm eq} \sim 1.6$~mG if $a=100$. However, 
the equipartition field is an upper limit, and  
the magnetic field in $V_{\rm x}$ is not necessarily the same
as that in the MERLIN emitter, as we will see in the next section.

\subsection{Synchrotron radio emission}

\begin{figure} 
\begin{center}
\includegraphics[angle=0,height=8.0cm]{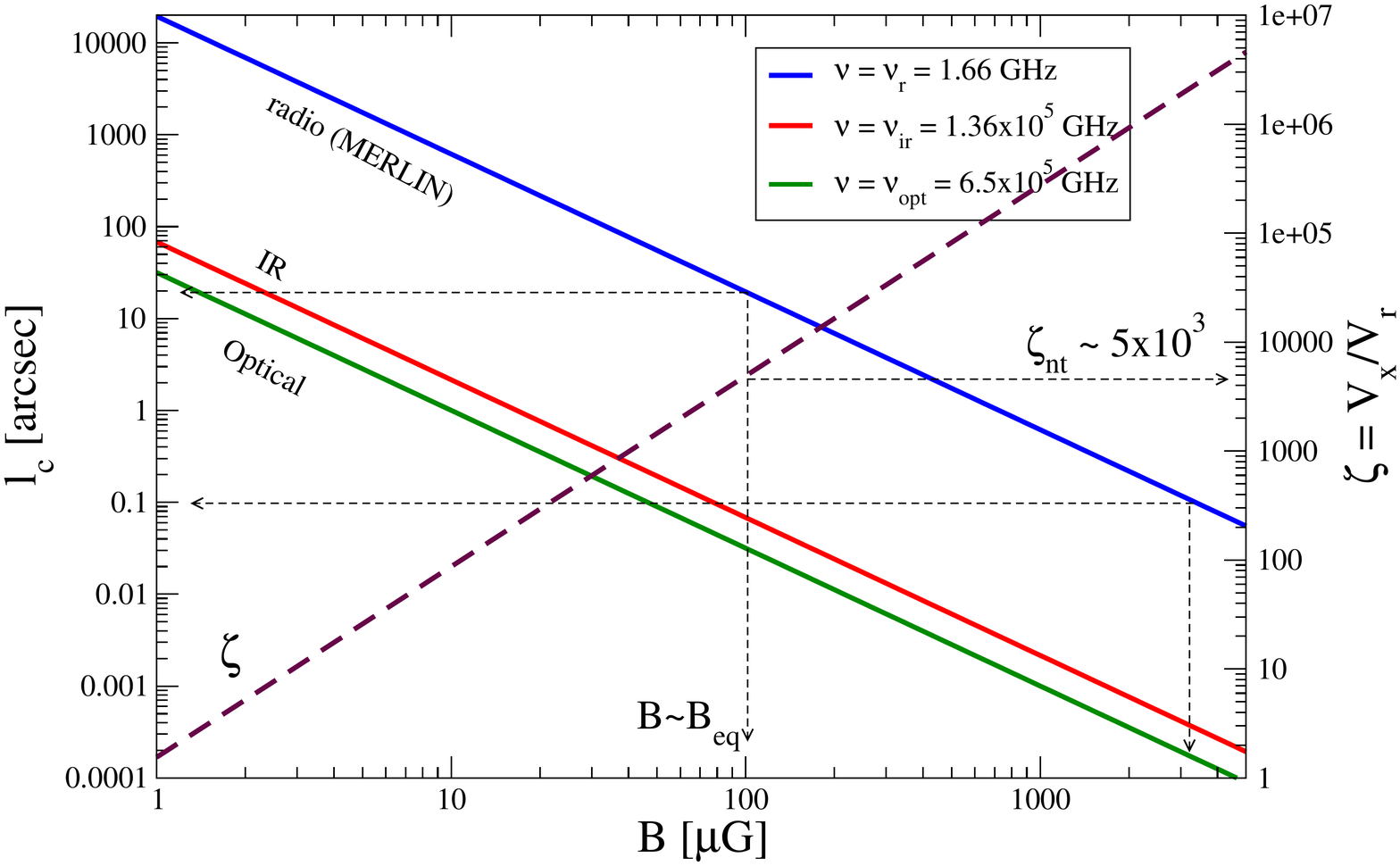}
\caption{\emph{Left axis} (solid lines): Synchrotron cooling length at 
radio (1.66~GHz), IR and optical frequencies.  
\emph{Right axis} (dashed line): X-rays to MERLIN emission volume.}
\label{losses}
\end{center}
\end{figure}

The synchrotron emission at $\nu_{\rm r}$ is produced by electrons with
$\gamma_{\rm r} \equiv \gamma(\nu_{\rm r})\sim 1.8\times10^3(B/100\,\mu{\rm G})^{-0.5}$, where  $\gamma(\nu) \sim 4.5\times10^{-4}(\nu/B)^{0.5}$
is the Lorentz factor of electrons emitting synchrotron radiation at frequency
$\nu$ in a magnetic field $B$. In a similar way to  $L_{\rm x}$, we write 
$L_{\rm r} \sim N_e(\gamma_{\rm r}) \gamma_{\rm r}^3/t_{\rm s}(\gamma_{\rm r}) V_{\rm r}$, where $t_{\rm s}=$7.5$\times$10$^8/(B^2\gamma)$ is the synchrotron cooling timescale and $V_{\rm r}$ is the  volume of the MERLIN emitter. 
Therefore, 
$L_{\rm x}/L_{\rm r} \sim (\gamma_{\rm x}/\gamma_{\rm r})^{3-p}(U_{\rm cmb}/U_{\rm mag})
\zeta$, where $U_{\rm mag} = B^2/(8\pi)$ and $\zeta \equiv V_{\rm x}/V_{\rm r}$,
and we find that 
\begin{equation}
\zeta \sim 4.9\times10^3
\left(\frac{B}{100\,{\rm \mu G}}\right)^{1.75}.
\end{equation}
In Fig.~\ref{losses} (right axis, dashed line) we can see that
$B \sim B_{\rm eq}$ corresponds to $\zeta_{\rm eq} =$5$\times$10$^3$.
Such a large ratio between emitting volumes is not implausible provided the 
magnetic field is
inhomogeneous in the shock downstream region and the  synchrotron 
emitter consists of features smaller than the MERLIN point spread function
(FWHM $0.15^{\prime\prime}$) as seen in parts of the MERLIN data. (Note
that if $V_{\rm x}$=$V_{\rm r}$, a very small magnetic field ($<\mu$G) would 
be needed to explain the observed fluxes.) Therefore,
X-ray emission is produced in volume $V_{\rm x} \gg V_{\rm r}$, as we can see
in Fig.~\ref{sketch},
and the magnetic field in $V_{\rm r}$ is much larger than the jet magnetic field
of the order of $\mu$G (Hardcastle \& Krause 2014). 
As we will discuss in the following, this may be the result of 
magnetic field amplification.

\section{The (synchrotron) hotspot as a magnetic field damping region}

The synchrotron ($l_{\rm s}$) and IC ($l_{\rm ic}$) cooling length of 
electrons with Lorentz factor 
$\gamma$  is $l_{\rm s,ic}(\gamma) = t_{\rm s,ic}(\gamma) v_{\rm sh}/7$, where 
$v_{\rm sh}/7$ is the velocity of the plasma downstream of the shock. 
The shock velocity is approximately the same as the jet velocity which we 
take to be $v_{\rm sh} = c/3$ (Steenbrugge \& Blundell, 2008).
We use $7$ as the shock compression ratio 
for a  non-relativistic shock whose downstream thermal pressure is 
dominated by relativistic electrons, although $4$ may 
still apply if non-relativistic ions dominate the pressure downstream of the
shock. Our conclusions
are not sensitive to the exact value of the shock compression ratio.

\subsection{X-ray emitter determined by adiabatic expansion}

The IC cooling lenght of X-ray emitting electrons is 
$l_{\rm ic}(\gamma_{\rm x}) \sim 10^{4} (3v_{\rm sh}/c)$~arcsec,
much larger than $l_{\rm x}$, and therefore
IC emission is not the mechanism that determines the size of the X-ray emitter. 
The synchrotron cooling length $l_{\rm s}$ of $\gamma_{\rm x}$-electrons 
is also greater than $l_{\rm x}$, unless the magnetic 
field in the X-ray emitting region is $\sim 360$~$\mu$G, greater than 
$B_{\rm eq}$ (unless $V_{\rm x} \sim 60$~arcsec$^3$). However, such a large
value of $B$ in  $V_{\rm x}$ would produce a synchrotron flux much greater
than that detected by the Very Large Array in the A-configuration
(Erlund et al. 2007).
Therefore, adiabatic expansion is probably the dominant cooling mechanism 
as the particles flow out of the hotspot.

\subsection{MERLIN emitter determined by magnetic field damping}
\label{damping}

The cooling length of electrons emitting synchrotron radiation at frequency
$\nu$ in a magnetic field $B$ is
\begin{equation}
\frac{l_{\rm s}(\nu)}{[\prime\prime]} \sim 12
\left(\frac{\nu}{\rm GHz}\right)^{-0.5} 
\left(\frac{B}{100\,{\rm \mu G}}\right)^{-1.5}
\left(\frac{v_{\rm sh}}{c/3}\right).
\label{l_cool}
\end{equation}
In the Fig.~\ref{losses} (left axis, solid lines)
we plot $l_{\rm s}(\nu_{\rm r})$, $l_{\rm s}(\nu_{\rm ir})$ and
$l_{\rm s}(\nu_{\rm opt})$. 
Optical and IR emission are almost co-spatial, with
$l_{\rm s}(\nu_{\rm ir})\sim l_{\rm s}(\nu_{\rm opt}) \sim 0.03^{\prime\prime}(B/100\,{\rm \mu G})^{-1.5}(3v_{\rm sh}/c)$ and indicating that these particles 
radiate most of their energy within $l_{\rm r}$\footnote{The diffuse IR 
and optical emission detected also in hotspots in other sources
has been suggested to be the result of reaceleration of non-thermal
electrons by second order Fermi acceleration (Brunetti et al. 2003). However,
this diffuse emission can be also the result of CMB photons up-scattered by 
electrons with $\gamma \sim 50$, or synchrotron emission
of electrons with $\gamma \sim \gamma_{\rm ir}\sim \gamma_{\rm opt}$ in a 
region with a smaller magnetic field, outside the MERLIN emitter.}. 
On the other hand, 
the synchrotron cooling length  of MERLIN emitting electrons is
$l_{\rm s}(\nu_{\rm r}) \sim 9.3^{\prime\prime}(B/100\,{\rm \mu G})^{-1.5}(3v_{\rm sh}/c)
\gg l_{\rm r}$. Even worse, the  
real hotspot extent downstream of the shock is $L < l_{\rm r}$ if
the jet is lying at an angle $\theta_{\rm j}<90^{\circ}$
with the line of sight. In particular, 
$L = (l_{\rm r} - D\cos\theta_{\rm j})/\sin\theta_{\rm j} \sim 0.1^{\prime\prime}$
when the hotspot is modelled as a cylinder of width $L$ and diameter 
$D = 3^{\prime\prime}$, and $\theta_{\rm j} = 73^{\circ}$ and 
$l_{\rm r}\sim 1^{\prime\prime}$ (see the right panel of Fig.~\ref{sketch}). 
In such a case, a very large magnetic field 
$B_{\rm cool} \sim 2.4 \,(3v_{\rm sh}/c)^{2/3}$~mG 
would be required to  match 
$l_{\rm s}(\nu_{\rm r}) = 0.1^{\prime\prime}$. This value is greater 
than $B_{\rm eq}$ for a wide range of $\gamma_{\rm min}$- and
$V_{\rm x}$-values, suggesting that the downstream extent of the compact 
emission 
detected at $\nu_{\rm r}$ is not the result of fast synchrotron cooling.
We suggest that the MERLIN emission region is determined by 
magnetic field amplification, as we explain below.

\section{Magnetic field amplification in mildly relativistic shocks}

Remarkable advances have been made in the last decade 
concerning DSA in the non- and ultra-relativistic regimes. In the former case,
the realisation that Non Resonant Hybrid instabilities (Bell 2004) 
in supernova remnants are fast 
enough to amplify (and maintain) the magnetic field by orders 
of magnitude (Vink \& Laming 2003) and accelerate
particles up to the knee ($10^{15.5}$~eV) of the cosmic ray spectrum, has
shed light on the origin of Galactic cosmic rays. 
In the ultra-relativistic case, however, theoretical studies show that Weibel 
instabilities  amplify the magnetic field on a short scale length, producing a 
rapid decay of the fluctuations and thereby
inhibiting particle acceleration to ultra high energies 
(Sironi et al. 2013, Reville \& Bell 2014). The mildly relativistic regime
has not been well studied (see however Brett et al. 2013).

In the present contribution we show a mildly 
relativistic ($v_{\rm sh}\sim c/3$) case study  where the synchrotron
cooling cannot determine the thickness of the radio emission detected with 
the high resolution
intereferometer MERLIN. Therefore, we suggest that the 
magnetic field  in the southern arc in 4C~74.26 is amplified
since $B \sim 100$~$\mu$G (required to explain the observations) 
is much larger than the 
expected value in the jet upstream of the termination shock (e.g. Hardcastle
\& Krause 2014). In addition to the thickness of the synchrotron emitter, the
cut-off of the synchrotron spectrum can give us an extra 
piece of information about the magnetic field.

\subsection{Synchrotron cut-off and magnetic field damping}

The synchrotron turnover $\nu_{\rm c}$ between $\nu_{\rm ir}$ and $\nu_{\rm opt}$ 
indicates that the maximum energy of non-thermal electrons is
\begin{equation}
E_{\rm c} = \gamma_{\rm c}\,m_ec^2 \sim 0.9
\left(\frac{\nu_{\rm c}}{\nu_{\rm opt}}\right)^{0.5}
\left(\frac{B}{100\,\mu{\rm G}}\right)^{-0.5}\,\,{\rm TeV},
\label{Ec}
\end{equation}
where $\gamma_{\rm c}\equiv \gamma_{\rm s}(\nu_{\rm c})$.
The standard assumption is that $\gamma_{\rm c}$ is
determined by a competition between shock acceleration and synchrotron cooling.
By equating $t_{\rm acc}(\gamma_{\rm c}) = t_{\rm s}(\gamma_{\rm c})$, where 
$t_{\rm acc} \sim 20 D/v_{\rm sh}^2$ is the acceleration timescale for the
case of a parallel shock, we find that
the electron diffusion coefficient $D$
is much larger than the Bohm value $D_{\rm Bohm}$:  
\begin{equation}
\frac{D}{D_{\rm Bohm}} \sim 10^6 \left(\frac{v_{\rm sh}}{c/3}
\right)^2
\left(\frac{\nu_{\rm c}}{\nu_{\rm opt}}\right)^{-1}, 
\end{equation}
and independent of $B$. Such a large diffusion coefficient 
is allowed if $B$ is structured on a scale $s$ much smaller than the
Larmor radius of the electrons being accelerated, producing 
$D\sim (r_{\rm g}/s) D_{\rm Bohm}$ and 
\begin{equation}
s \sim 2\times10^{7} \left(\frac{\nu_{\rm c}}{\nu_{\rm opt}}\right)^{1.5}
\left(\frac{B}{100\,{\rm \mu G}}\right)^{-1.5}
\left(\frac{v_{\rm sh}}{c/3}\right)^{-2} \,\,{\rm cm}.
\label{lambda}
\end{equation}
In comparison the ion skin-depth is 
$c/\omega_{\rm pi} \sim 2.3 \times10^9 \,(n/10^{-4}\,{\rm cm^{-3}})^{-0.5}$~cm,
where $n$ is the particle density downstream of the shock ($n = 7 n_{\rm j}$, 
where the $n_{\rm j}$ is the jet density),   and 
\begin{equation}
\frac{s}{c/\omega_{\rm pi}} \sim 0.01 
\left(\frac{\nu_{\rm c}}{\nu_{\rm opt}}\right)^{1.5}
\left(\frac{v_{\rm sh}}{c/3}\right)^{-2}
\left(\frac{B}{100\,{\rm \mu G}}\right)^{-1.5}
\left(\frac{n}{10^{-4}\,{\rm cm^{-3}}}\right)^{0.5}.
\end{equation}
Therefore, 
\begin{equation}
B \le 4.6 
\left(\frac{\nu_{\rm c}}{\nu_{\rm opt}}\right)
\left(\frac{v_{\rm sh}}{c/3}\right)^{-4/3}
\left(\frac{n}{10^{-4}\,{\rm cm^{-3}}}\right)^{3}\,{\rm \mu G}
\end{equation}
is required to satisfy the condition
$s \ge c/\omega_{\rm pi}$ in the case that $E_{\rm c}$ is constrained by
synchrotron cooling. Note however that $B \propto n^3\propto n_{\rm j}^3$.
In Araudo et al. (2016, submitted) we explore this condition in depth.

In the case that the magnetic field is amplified by the Weibel 
instability, small-scale turbulence ($s \sim c/\omega_{\rm pi}$) scatters 
non-thermal electrons during DSA (Sironi et al. 2011).  However, 
the hotspot magnetic field has to survive over distances 
$\sim$0.2~kpc ($\sim$0.1$^{\prime\prime}$) downstream of the jet reverse shock, 
which are much larger than those predicted by numerical simulations.
The same discrepancy was found in Gamma-Ray Bursts 
(e.g. Medvedev et al. 2005, Pe'er \& Zhang, 2006).  
Chang et al. (2008) have shown that the magnetic field
generated by Weibel instabilities in ultra relativistic plasmas is 
maintained constant over a distance
$\sim 100\,c/\omega_{\rm p}$ downstream of the shock, and then 
decays as $\propto l^{-1}$, where $l$ is the distance downstream of the shock.
(See also Lemoine 2015 for a similar study in the non-linear regime.)

\section{Conclusions}
\label{discussion}

We model the radio to X-ray emission  in the 
southern hotspot of the FR~II radiogalaxy 4C74.26. 
Our study is based on three  key features:
\begin{enumerate}
\item The MERLIN emission region is too thin to be the result of
fast synchrotron cooling. 
\item The radio to IR spectrum ($\alpha = 0.75$) is too flat for the emitting
volume to be determined 
by synchrotron cooling through this wavelength range.
\item The turnover of the synchrotron spectrum at IR/optical frequencies 
requires $D \gg D_{\rm Bohm}$ for any reasonable shock velocity.
\end{enumerate}
These three features fit well  in a scenario in which the MERLIN radio 
emission traces out the region where the magnetic field is amplified
by plasma instabilities with small length scale.
The magnetic field decays quickly behind the shock accounting 
for the maximum energy  of accelerated electrons at $E_{\rm c} \sim$~TeV.
These electrons continue up-scattering CMB photons, thus producing IC
X-ray emission downstream of the shock after the MERLIN
radio emission has ceased. 

The magnetic field 
in equipartition with non-thermal electrons in the MERLIN  emission region 
is  $\sim$100~$\mu$G  
and similar to the values obtained by other authors. 
An unrealistically large magnetic field
$B_{\rm cool}\sim 2.4\,(3v_{\rm sh}/c)^{2/3}$~mG would be needed to
explain the compact radio emission in terms of synchrotron cooling.  
If $B \sim 100$~$\mu$G  in the synchrotron  emission region,
the maximum energy of non-thermal electrons is $\sim$~TeV, 
(Eq.~\ref{Ec}). If ions are accelerated 
as well, protons with energy $\sim$~TeV diffuse also with 
$D \gg D_{\rm Bohm}$ and 
then the maximum proton energy at the termination shock of 4C74.26
is only $100$~TeV instead of the $100$~EeV indicated by the Hillas parameter.
This may have important implications for the understanding of the origins 
of ultra high energy cosmic rays.

\acknowledgments 
We thank the referee for a constructive report.
A.T.A. and A.R.B. thank the organisers of the HEPRO~V conference for their 
kind  hospitality. We acknowledge support from
the European Research Council under the European
Community's Seventh Framework Programme (FP7/2007-2013)/ERC grant agreement 
no. 247039, and from the UK
Science and Technology Facilities Council under grant No. ST/K00106X/1.


\begin{references}

\reference Araudo, A.~T., Bell, A.~R., Blundell, K.~M. 2015, \apj, 806, 243
\reference Araudo, A.~T., Bell, A.~R., Crilly, A., Blundell, K.~M. 2016
(submitted) 
\reference  Bell, A.~R. 1978, \mnras, 182, 147
\reference  Bell, A.~R. 1978, \mnras, 182, 444
\reference  Bell, A.~R. 2004, \mnras, 353, 550
\reference Bret, A., Stockem, A., Fiuza, F., Ruyer, C., Gremillet, L., 
Narayan, R., Silva, L.O. 2013, Physics of Plasmas, 20, 042102
\reference Brunetti, G.; Mack, K.-H.; Prieto, M. A.; Varano, S. 2003, \mnras, 345, 40L
\reference Chang, P., Spitkovsky, A., Arons, J. 2008, \apj, 374, 378
\reference Erlund, M. C., Fabian, A. C., Blundell, K. M., Moss, C., 
Ballantyne, D. R. 2007, \mnras, 379, 498
\reference Erlund, M.~C., Fabian, A.~C., Blundell, K.~M., Crawford, C.~S., 
Hirst, P. 2010, \mnras, 404, 629
\reference Hardcastle, M. J.; Harris, D. E.; Worrall, D. M.; Birkinshaw, M.
2004, \mnras, 612, 729
\reference Hardcastle, M. J., Krause, M.G.H. 2014, \mnras, 443, 1482
\reference Lemoine, M. 2015, Journal of Plasma Physics, 81, 455810101
\reference Meisenheimer, K., Roser, H.-J., Hiltner, P.~R., Yates, M.~G., 
Longair, M.~S., Chini, R., Perley, R.~A. 1989, \aap, 219, 63
\reference Pe'er, A., Zhang, B. 2006, \apj, 653, 454
\reference Reville, B., Bell, A. R. 2014, \mnras, 439, 2050
\reference Ryter, Ch. E. 1996, \apss, 236, 285
\reference Sironi, L., Spitkovsky, A. 2011, \apj, 726, 75
\reference Sironi, L., Spitkovsky, A., Arons J. 2013, \apj, 771, 54
\reference Steenbrugge, K.~C., Blundell, K.~M. 2008, \mnras, 388, 1457
\reference Vink, J., Laming, J.M. 2003, \apj, 554, 758
\reference Werner, M.~W., Murphy, D.~W., Livingston, J.~H., Gorjian, V., Jones, D.~L.; Meier, D.~L.; Lawrence, C.~R. 2012, \apj, 759, 86
\end{references}
\end{document}